\documentstyle[a4,12pt]{article}
\begin{document}
\pagestyle{empty}
\begin{titlepage}
\rightline{IC/94/274}
\rightline{UTS-DFT-94-10}
\vspace{1.5cm}
\begin{center}
\begin{Large}
{\bf Model independent constraints on anomalous gauge boson self-couplings from
$e^+e^-$ colliders with longitudinally polarized beams}

\end{Large}

\vspace{2cm}

{\large  A. A. Pankov\hskip 2pt\footnote{
E-mail: PANKOV\%GPI.GOMEL.BY@RELAY.USSR.EU.NET}
}\\[0.5cm]
{\it Gomel Polytechnical Institute, Gomel, 246746 Belarus, CIS}\\[1cm]
\vspace{3mm}
{\large  N. Paver\hskip 2pt\footnote{Also supported by the Italian
Ministry of University, Scientific Research and Technology (MURST).}
}\\[0.5cm]
{\it Dipartimento di Fisica Teorica, Universit\`{a} di Trieste, 34100
Trieste, Italy}
\\
{\it Istituto Nazionale di Fisica Nucleare, Sezione di Trieste, 34127 Trieste,
Italy}\\[1cm]
\end{center}

\vspace{1.5cm}

\begin{abstract}
\noindent We explore the possibility of deriving model independent limits on
the anomalous trilinear electroweak gauge boson couplings from high energy
$e^+e^-\to W^+W^-$, by combining the cross sections for the different initial
and final states polarizations integrated with suitable kinematical cuts.
In the case of the CP conserving couplings the limits can be disentangled,
and are given by simple mathematical expressions. Numerical results show the
advantages of this approach, in particular the important role of polarization
in improving the bounds.
\end{abstract}
\end{titlepage}

\pagestyle{plain}
\noindent
The precise measurement of the $WW\gamma$ and $WWZ$ couplings
is essential for the confirmation of the non abelian gauge structure of the
Standard Model (SM). In this regard, a special role is played by the process
\begin{equation}e^++e^-\to W^++W^-\label{proc}\end{equation}
at the planned high energy $e^+e^-$ colliders, because in this case deviations
from the SM are significantly enhanced by increasing the CM energy, and
correspondingly the sensitivity is improved.
In general, the trilinear gauge boson interaction includes CP violating
couplings as well as CP conserving ones. The set of measurements sensitive to
the CP violating couplings and their separation was discussed in
Ref.\cite{gounaris1}. Furthermore, the possibility of separately constraining
the C and P violating (but CP conserving) anapole coupling, using process
(\ref{proc}) with initial beams longitudinal polarization, was discussed in
Ref.\cite{anom2}.
Therefore, we shall limit here to the derivation of constraints for
the CP conserving couplings which appear in the effective Lagrangian
\cite{hagiwara,gounaris2}
\begin{eqnarray}{\cal L}_1&=&-ie\hskip 2pt\left[A_\mu\left(W^{-\mu\nu}W^+_\nu-
W^{+\mu\nu}W^-_\nu\right)+F_{\mu\nu}W^{+\mu}W^{-\nu}\right]-ie\hskip 2pt
x_{\gamma}\hskip 2pt F_{\mu\nu}W^{+\mu}W^{-\nu}
\nonumber\\
&-& ie\hskip 2pt \left(\cot\theta_W+\delta_Z\right)\hskip 2pt \left[Z_\mu
\left(W^{-\mu\nu}W^+_\nu-W^{+\mu\nu}W^-_\nu\right)+Z_{\mu\nu}W^{+\mu}W^{-\nu}
\right]\nonumber\\
&-&ie\hskip 2pt x_Z\hskip 2pt Z_{\mu\nu}W^{+\mu}W^{-\nu}
 +ie\hskip 2pt\frac{y_{\gamma}}{M_W^2}\hskip 2pt F^{\nu\lambda}W^-_{\lambda\mu}
W^{+\mu}_{\ \ \nu}+ie\hskip 2pt\frac{y_Z}{M_W^2}\hskip 2pt
Z^{\nu\lambda}
W^-_{\lambda\mu}W^{+\mu}_{\ \ \nu}\hskip 2pt ,\label{lagra}\end{eqnarray}
where $W_{\mu\nu}^{\pm}=\partial_{\mu}W_{\nu}^{\pm}-
\partial_{\nu}W_{\mu}^{\pm}$ and ${Z_{\mu\nu}=\partial_{\mu}Z_{\nu}-
\partial_{\nu}Z_{\mu}}$.
According to Eq.(\ref{lagra}), in general we have five independent couplings,
with SM values $\delta_Z=x_{\gamma}=x_Z=y_{\gamma}=y_Z=0$. Since the
unpolarized cross section depends on all coupling constants, it should be
difficult to separately constrain them using this observable only. To
disentangle and limit the couplings in a model independent way one would
need more information. This should be provided by the separate measurements
of the cross
sections for initial and final states polarizations, which depend on
independent combinations of the coupling constants. Ideally, the three
possible $W^+W^-$ polarizations ($LL$, $TL$ and $TT$), combined with the
two longitudinal $e^-\hskip 2pt e^+$ ones ($RL$ and $LR$) would
determine a sufficient set of observable cross sections. The purpose of this
note is to illustrate the role of polarizations to derive model independent
bounds on the five anomalous couplings and to quantitatively assess the
corresponding expected sensitivities.\par
The basic objects are the deviations of the polarized cross sections from the
SM values
\begin{equation}\Delta\sigma=\sigma-\sigma_{SM},\label{deltasig}\end{equation}
where, in terms of the Born $\gamma$-, $Z$- and $\nu$-exchange amplitudes and
their deviations from the SM expressions due to the anomalous gauge couplings:
\begin{eqnarray}d\sigma&\propto&
\vert{\cal A}(\gamma)+\Delta{\cal A}(\gamma)+
{\cal A}(Z)+\Delta{\cal A}(Z)+{\cal A}_1(\nu)\vert^2+
\vert{\cal A}_2(\nu)\vert^2,\nonumber \\
d\sigma_{SM}&\propto&
\vert {\cal A}(\gamma)+{\cal A}(Z)+{\cal A}_1(\nu)\vert^2
+\vert{\cal A}_2(\nu)\vert^2.\label{deltaa}\end{eqnarray}
In Eq.(\ref{deltaa}) we have distinguished the
$\nu$- exchange amplitudes with $\vert\lambda-\bar\lambda\vert\leq 1$ and
$\vert\lambda-\bar\lambda\vert=2$, where $\lambda$ and $\bar\lambda$ are
the $W^-$ and $W^+$ helicities. With the aid of explicit formulae for the
helicity amplitudes given, {\it e.g.}, in Ref.\cite{gounaris2}, one
easily finds for the specific initial and final polarizations the following
dependence of the amplitudes deviations $\Delta{\cal A}$'s in
Eq.(\ref{deltaa}):
\begin{eqnarray} \Delta{\cal A}_{LL}^{ab}(\gamma)&\propto&x_{\gamma}
\nonumber\\
\Delta{\cal A}_{LL}^{ab}(Z)&\propto&\left(x_Z+\delta_Z\hskip 1pt
\frac{3-\beta_W^2}{2}\right)\hskip 1pt g_e^a,\label{deltall}\end{eqnarray}
\begin{eqnarray} \Delta{\cal A}_{TL}^{ab}(\gamma)&\propto&x_{\gamma}+y_{\gamma}
\nonumber\\
\Delta{\cal A}_{TL}^{ab}(Z)&\propto&\left(x_Z+y_Z+2\hskip 1pt\delta_Z\right)
\hskip 1pt g_e^a,\label{deltatl}\end{eqnarray}
and
\begin{eqnarray} \Delta{\cal A}_{TT}^{ab}(\gamma)&\propto&y_{\gamma}
\nonumber\\
\Delta{\cal A}_{TT}^{ab}(Z)&\propto&\left(y_Z+\delta_Z\hskip 1pt
\frac{1-\beta_W^2}{2}\right)\hskip 1pt g_e^a.\label{deltatt}\end{eqnarray}
In Eqs.(\ref{deltall})-(\ref{deltatt}):
${\beta_W=\sqrt{1-4M_W^2/s}}$, the lower indices $LL$, $TL$ and $TT$ refer to
the final $W^-W^+$ polarizations, the upper indices $a$ and $b$ indicate the
initial $e^-$ $e^+$ $RL$ or $LR$ polarizations, and
$g_e^R=\tan\theta_W$ and $g_e^L=g_e^R\left(1-1/2\sin^2\theta_W\right)$
are the corresponding electron couplings. One
should notice that $\sigma_{LL}$, $\sigma_{TL}$ and $\sigma_{TT}$ depend on
the combinations ($x_{\gamma},x_Z+\delta_Z(3-\beta_W^2)/2$),
($x_{\gamma}+y_{\gamma},x_Z+y_Z+2\delta_Z$) and
($y_{\gamma},y_Z+\delta_Z(1-\beta_W^2)/2$) respectively.\par
In order to assess the sensitivity of the different cross sections to the
gauge boson couplings, we define a $\chi^2$ function as
\begin{equation}\chi^2=\left(\frac{\Delta\sigma}{\delta\sigma_{SM}}\right)^2,
\label{chisquare}\end{equation}
where $\sigma\equiv\sigma(z_1,z_2)=
\int_{z_1}^{z_2}\left(d\sigma/dz\right)dz$ with
$z=\cos\theta$ and $\delta\sigma_{SM}$ is the accuracy experimentally
obtainable on $\sigma(z_1,z_2)_{SM}$. Including both statistical and
systematical errors,
${\delta\sigma_{SM}=\sqrt{(\delta\sigma_{stat})^2+(\delta\sigma_{syst})^2}}$,
where $(\delta\sigma/\sigma)_{stat}=1/\sqrt{L_{int}\varepsilon_W\sigma_{SM}}$,
with  $L_{int}$ the integrated luminosity and $\varepsilon_W$ the
efficiency for $W^+W^-$ reconstruction in the considered polarization state.
For that we take the channel of lepton pairs ($e\nu+e\mu$) plus two hadronic
jets, which corresponds to $\varepsilon_W\simeq 0.3$
\cite{frank}-\cite{anlauf}.\footnote{Actually, this reconstruction efficiency
might be somewhat smaller, depending on the detector \cite{frank}. On the
other hand, for our estimates we have taken a rather conservative choice for
the integrated luminosity, while
recent progress in machine design seems to indicate that quite larger
values are attainable \cite{settles} and can compensate for the reduction of
$\varepsilon_W$.} Then, as a criterion to
derive allowed regions of the coupling constants, we will impose that
$\chi^2\leq\chi^2_{crit}$, where $\chi^2_{crit}$ is a number which specifies
a chosen confidence level and in principle can depend on the details of the
analysis. In this procedure, an essential role is played by the values of
$z_1$ and $z_2$. Indeed, for each initial and final polarizations, it is
possible to choose the upper and lower integration limits in such a way
as to get maximum sensitivity of the corresponding polarized cross sections to
the combinations of the coupling constants in
Eqs.(\ref{deltall})-(\ref{deltatt}) \cite{frank, anom2}. The search of these
`optimal' integration regions can be done numerically, by plotting in each case
the $\chi^2$ function (\ref{chisquare}) {\it vs.} the anomalous couplings for
different $z_1$ and $z_2$, and by looking for the values of $z_1$ and $z_2$
which minimize the range of couplings such the inequality
$\chi^2\leq\chi^2_{crit}$ holds.
To be closer to a possible experimental situation, we have taken into
account that in practice the cross section should be
\begin{equation}\sigma=\frac{1}{4}\left[(1+P_1)\cdot (1-P_2)\hskip 2pt
\sigma^{RL}+(1-P_1)\cdot (1+P_2)\hskip 2pt\sigma^{LR}\right],\label{gcr}
\end{equation}
where $P_1$ ($P_2$) are less than unity, and represent the actual degrees of
longitudinal polarization of $e^-$ ($e^+$). In the sequel we shall consider
as $RL$ or $LR$ the simplified situations $P_1=-P_2=P>0$ and $P_1=-P_2=-P$,
respectively, with $P=0.9$ as a possible value \cite{prescott}.\par
In Fig.1 we show an example relevant to the cross
sections for unpolarized $W$'s and both unpolarized and polarized electrons.
For simplicity only the coupling $x_{\gamma}$ is considered, with all the
other ones taken equal to their SM values. The inputs as well as the resulting
optimal kinematical regions are presented in the caption of the figure.
The allowed limits on the values of $x_{\gamma}$ are at the
two standard deviations level (or 95\% CL), which for our analysis corresponds
to $\chi^2_{crit}=4$. In this example, as well as in the following analysis,
we have taken $(\delta\sigma_/\sigma)_{syst}=2\%$ as currently assumed
\cite{frank}. The role of optimal kinematics and of longitudinal initial
polarizations is particularly evident in this particular example. This is
connected to the fact that for unpolarized and $LR$ $e^-\hskip 1pt e^+$ the
relevant angular distribution of $\Delta\sigma$ in the numerator of
Eq.(\ref{chisquare}) has a zero, so that the integration over the whole
angular range allowed by an  experimental $10^\circ$ cut
($z_1=-0.98$, $z_2=0.98$) would lead to a reduced signal from the anomalous
coupling. Furthermore, the cross section for final $TT$ and unpolarized
$W^+W^-$ and any initial polarization includes the contribution of the
$\nu$-mediated amplitudes with $\lambda-{\bar\lambda}=\pm 2$
(see Eq.(\ref{deltaa})), which by far dominates in the forward direction and
thus strongly suppresses the signal.\par
The general situation regarding the optimal $z_1$ and $z_2$ for the
various cross sections, and the corresponding statistical uncertainties, is
presented in Table 1 for two values of the CM energy and the planned
luminosities \cite{settles, treille}.
\begin{table}
\centering
\begin{tabular}{|c|c|c|c|}
\hline
$$ & $e^-_Re^+_L\to W^-_LW^+_L$ & $e^-_Re^+_L\to W^-_LW^+_T+ W^-_TW^+_L$
& $e^-_Re^+_L\to W^-_TW^+_T$\\ \hline
$z_{opt}$ &$0.98\hskip 2pt(0.98)$ & $0.98\hskip 2pt(0.98)$
&$0.22\hskip 2pt(0.22)$  \\ \hline
$\sigma_{SM}(fb)$ & $76\hskip 2pt(19)$ & $28\hskip 2pt(1.9)$ &
$2.9\hskip 2pt(0.6)$  \\ \hline
$\delta_{stat}(\%)$ & $4.7\hskip 2pt(5.9)$ & $7.6\hskip 2pt(18.6)$ &
$24\hskip 2pt (32)$  \\ \hline\hline
$$ & $e^-_Le^+_R\to W^-_LW^+_L$ & $e^-_Le^+_R\to W^-_LW^+_T+ W^-_TW^+_L$
& $e^-_Le^+_R\to W^-_TW^+_T$ \\ \hline
$z_{opt}$ &$0.85\hskip 2pt(0.96)$ & $-0.35\hskip 2pt(0.98)$
&$0.13\hskip 2pt(0.13)$  \\ \hline
$\sigma_{SM}(fb)$ & $342\hskip 2pt(87)$ & $44\hskip 2pt(35)$ &
$780\hskip 2pt(187)$  \\ \hline
$\delta_{stat}(\%)$ & $2.2\hskip 2pt(2.8)$ & $6.2\hskip 2pt(4.4)$ &
$1.5\hskip 2pt (1.9)$  \\ \hline
\end{tabular}
\caption{Optimal integration regions for $E_{CM}=0.5\hskip 2pt TeV$ and
$1\hskip 2pt TeV$ (in parentheses). Integrated luminosities
$L_{int}= 20\hskip 2pt fb^{-1}$ and $50\hskip 3pt fb^{-1}$ respectively;
$P_1=-P_2=0.9$ ($RL$), $P_1=-P_2=-0.9$ ($LR$).}\label{tab:tab1}
\end{table}
It turns out that in all cases one can take for
the lower integration limit the minimum allowed value $z_1=-0.98$. In fact, at
this point the relative deviation $\Delta\sigma/\sigma_{SM}$ and
$\Delta\sigma/\delta\sigma_{SM}$ are both maximal and correspondingly so is
the sensitivity to the anomalous couplings. This
reflects the fact that the `background' $\nu$-exchange contribution to the
cross section is minimal in the backward direction. Consequently, the
searched for optimal kinematical region can be specified by only
$z_{opt}\equiv z_2$.\par
Applying the procedure outlined above to the reaction $e^-e^+\to W^-_LW^+_L$,
and taking into account the results of Table 1, we obtain the
$\chi^2=4$ contours allowed to the combinations of couplings of
Eq.(\ref{deltall}) by each initial polarization. These are represented for
$E_{CM}=500\hskip 2pt GeV$ in Fig.2. The allowed regions enclosed by those
contours are all elliptical (the $RL$ and $LR$ ones are extremely flattened,
depending on $P_1$ and $P_2$, and therefore only their parts relevant to
the intersections are drawn in Fig.2). Of the four common intersections, whose
existence for $RL$ and $LR$ initial polarizations
is assured by $g_e^L\simeq -g_e^R$, only one includes the region
around the SM values of the trilinear gauge boson couplings. One finds
analytically that the position of the
intersections does not depend on the polarizations $P_1$ and $P_2$, so that
the only way to exclude the three intersections not containing the
SM point would be to change the CM energy.\par
Concentrating on the region around the origin, in Fig.3 we represent a
magnification of Fig.2 and the area allowed by the combination of the two
observables $\sigma^{RL}$ and $\sigma^{LR}$, taking $\chi^2_{crit}=5.9$ and
$E_{CM}=500\hskip 3pt GeV$ (the smaller region would be the result for
$E_{CM}=1\hskip 3pt TeV$). The area allowed by the unpolarized cross section
does not add any significant information, and is included in the figure
just for comparison.\par
{}From Fig.3 one can read the constraints, which can be expressed by the
following inequalities:
\begin{equation}-\alpha_1^{LL}<x_{\gamma}<\alpha_2^{LL},\label{alphall}
\end{equation}
\begin{equation}-\beta_1^{LL}<x_Z+\delta_Z\frac{3-\beta_W^2}{2}<\beta_2^{LL},
\label{betall}\end{equation}
where $\alpha_{1,2}^{LL}$ and $\beta_{1,2}^{LL}$ are the projections of the
combined allowed area on the horizontal and vertical axes, respectively,
and clearly depend on the inputs for energy, polarization, kinematics and
luminosity. One can notice that in the process $e^+e^-\to W^+_LW^-_L$ the
initial state polarization allows to bound $x_{\gamma}$ separately. The
typical bounds for the inputs in the caption of Table 1 are of the order of
$10^{-3}$, as can be seen from Fig.3. This order of magnitude is simply
explained by considering, {\it e.g.}, the amplitude relevant to
$\sigma_{LL}^{RL}$:
\begin{equation}{\cal A}^{RL}_{LL}=\frac{s}{M_W^2}\hskip 2pt
\left[\frac{3-\beta_W^2}{2}\left(1-\chi_Z\hskip 2pt\cot\theta_W g^R_e\right)+
x_{\gamma}-\chi_Z\hskip 2ptg^R_e\hskip 2pt\left(x_Z
+\delta_Z\frac{3-\beta_W^2}{2}\right)\right],\label{arl}\end{equation}
where $\chi_Z= s/\left(s-M_Z^2\right)$ is the $Z$ boson propagator. The cross
section is given by
\begin{equation}\frac{d\sigma^{RL}_{LL}}{dz}=\frac{\pi\alpha_{\it e.m.}^2
\beta_W^3}{8s}\hskip 2pt \left(1-z^2\right)
\vert{\cal A}^{RL}_{LL}\vert^2.\label{crsll}\end{equation}
{}From the requirement $\chi^2\leq\chi^2_{crit}=4$ one has for
$x_Z=\delta_Z=0$:
\begin{equation}\vert x_\gamma\vert\leq \frac{1}{4}\sqrt{\chi^2_{crit}}
\hskip 2pt\left(3-\beta^2_W\right)\hskip 2pt\left(1-\chi_Z\right)\hskip 2pt
\left(\frac{\delta\sigma_{SM}}{\sigma_{SM}}\right).\label{typical}
\end{equation}
{}From Table 1 and the assumed 2\% systematic error we have
$\left(\delta\sigma_{SM}/\sigma_{SM}\right)\simeq 5\%$ and from
(\ref{typical}): $\vert x_\gamma\vert\leq 1.8\times 10^{-3}$.\par
We now turn to the other polarized cross section, and repeat the same analysis
there. In Fig.4 we represent the analogous of Fig.3 for the combinations of
coupling constants in Eq.(\ref{deltatl}), which results from
$e^+e^-\to W^+_TW^-_L+W^+_LW^-_T$. In this case, one obtains the following
inequalities, analogous to (\ref{alphall}) and (\ref{betall}):
\begin{equation}-\alpha_1^{TL}<x_{\gamma}+y_{\gamma}<\alpha_2^{TL},
\label{alphatl}\end{equation}
\begin{equation}-\beta_1^{TL}<x_Z+y_Z+2\delta_Z<\beta_2^{TL}.
\label{betatl}\end{equation}
\par Finally, from $e^+e^-\to W^+_TW^-_T$ one obtains for the combinations of
Eq.(\ref{deltatt}) the allowed regions in Fig.5 and the corresponding
inequalities:
\begin{equation}-\alpha_1^{TT}<y_{\gamma}<\alpha_2^{TT},\label{alphatt}
\end{equation}
\begin{equation}-\beta_1^{TT}<y_Z+\frac{1-\beta_W^2}{2}\delta_Z<\beta_2^{TT}.
\label{betatt}\end{equation}
The less restrictive limits in Fig.5 are determined by the larger width of
the region enclosed by the $LR$ contours, mainly due to the
dominance in this channel of the $\vert\lambda-{\bar\lambda}\vert=2$
contribution which significantly reduces the sensitivity even in the optimal
kinematical region. Also, we can notice that with initial state polarization
$e^+e^-\to W^+_TW^-_T$ can constrain $y_{\gamma}$ separately.\par
Finally, from Eqs.(\ref{alphall}) to (\ref{betatt}) one can obtain the
simple, model independent and separate bounds:
\begin{equation}-\frac{1}{\beta_W^2}B_2<\delta_Z<\frac{1}{\beta_W^2}B_1,
\label{B1}\end{equation}
\begin{equation}-\left(\beta_1^{LL}+\frac{3-\beta_W^2}{2\beta_W^2}B_1\right)<
x_Z<\beta_2^{LL}
+\frac{3-\beta_W^2}{2\beta_W^2}B_2\label{B2},\end{equation}
\begin{equation}-\left(\beta_1^{TT}+\frac{1-\beta_W^2}{2\beta_W^2}B_1\right)
<y_Z<\beta_2^{TT}
+\frac{1-\beta_W^2}{2\beta_W^2}B_2,\label{B3}\end{equation}
where $B_1=\beta_1^{LL}+\beta_1^{TT}+\beta_2^{TL}$ and
$B_2=\beta_2^{LL}+\beta_2^{TT}+\beta_1^{TL}$. These constraints should be
joined with (\ref{alphall}) and (\ref{alphatt}) for $x_\gamma$ and
$y_\gamma$, respectively.\par
We have one more constraint from the combination of inequalities
(\ref{alphall}) and (\ref{alphatl}):
\begin{equation}-\left(\alpha_1^{TL}+\alpha_2^{LL}\right)<y_\gamma<
\alpha_1^{LL}+\alpha_2^{TL},\label{ygamma}\end{equation}
which has to be compared with (\ref{alphatt}). It turns out that
for $E_{CM}=500\hskip 3pt GeV$ the most stringent limitation for $y_\gamma$ is
determined by (\ref{ygamma}), whereas (\ref{alphatt}) is the most restrictive
one for $1\hskip 3pt TeV$.\par
The numerical results from these relations, and the chosen
inputs for the luminosity and the initial polarization, are collected in
Table 2.

\begin{table}
\centering
\begin{tabular}{|c|c|c|c|c|c|}
\hline
$E_{CM}(TeV)$ & $x_\gamma\hskip2pt (10^{-3})$ & $y_\gamma\hskip2pt (10^{-3})$
& $\delta_Z\hskip2pt (10^{-3})$ & $x_Z\hskip2pt (10^{-3})$
& $y_Z\hskip2pt (10^{-3})$ \\ \hline
$0.5$ & $-1.8\div 1.8$ & $-8.6\div 9.2$ &$-40\div 40$ & $-45\div 45$
& $-22\div 22$ \\ \hline
$1$ & $-0.5\div 0.5$ & $-3.0\div 3.0$ &$-13\div 13$ & $-14\div 14$
& $-5.7\div 6.0$ \\ \hline
\end{tabular}
\caption{Model independent limits on the non-standard gauge boson couplings at
the 95\% CL. Same inputs as in Table 1.}\label{tab:tab2}
\end{table}
\par It should be interesting to specialize the previous analysis to
`physically' motivated models, where nonstandard trilinear gauge boson
couplings originate from some new interaction acting at a higher scale
$\Lambda$ much greater than
the Fermi scale. A popular class of models assumes for such an
interaction an $SU(2)\times U(1)$ spontaneously broken local symmetry, with
gauge bosons $\gamma$, $W$ and $Z$ and one Higgs doublet
\cite{zeppenfeld}--\cite{gounaris3}.
Accordingly, the weak interaction Lagrangian should be given by the
combination
\begin{equation}{\cal L}_W={\cal L}_{SM}+\sum_d\sum_k
\frac{f^{(d)}_k}{\Lambda^{d-4}}{\cal O}^{(d)}_k, \label{lagra1}\end{equation}
where ${\cal L}_{SM}$ is the familiar, renormalizable SM Lagrangian, and
the gauge invariant effective operators ${\cal O}^{(d)}_k$ are ordered by
dimension $d$ and represent the low energy effect of the new interaction,
giving rise in particular to the anomalous gauge boson couplings. From the
good agreement of the measured fermion couplings with the SM ones, one assumes
that new contributions to these couplings can be neglected. Then, limiting to
dimension 6 operators, the relevant $C$ and $P$ conserving operators are
\cite{buch}
\begin{eqnarray}{\cal O}^{(6)}_{WWW}&=&Tr\left[{\hat W}_{\mu\nu}
{\hat W}^{\nu\rho}{\hat W}^{\mu}_{\rho}\right],\nonumber\\
{\cal O}^{(6)}_W&=&\left(D_\mu\Phi\right)^\dagger{\hat W}^{\mu\nu}
\left(D_\nu\Phi\right),\nonumber\\
{\cal O}^{(6)}_B&=&\left(D_\mu\Phi\right)^\dagger{\hat B}^{\mu\nu}
\left(D_\nu\Phi\right).\label{oper}\end{eqnarray}
Here, $\Phi$ is the Higgs doublet and, in terms of the $B$ and $W$
field strengths: ${\hat B}^{\mu\nu}=i(g^\prime/2)B^{\mu\nu}$,
${\hat W}^{\mu\nu}=i(g/2){\vec\tau}\cdot{\vec W}^{\mu\nu}$ with $\vec\tau$
the Pauli matrices. The contributions to the anomalous vector boson couplings
are:
\begin{equation}x_\gamma=\cos^2\theta_W\hskip 2pt
\left(f^{(6)}_B+f^{(6)}_W\right)\hskip 2pt\frac{M^2_Z}{2\Lambda^2};
\qquad y_\gamma=f^{(6)}_{WWW}
\hskip 2pt\frac{3M^2_Wg^2}{2\Lambda^2}; \label{deltaz}\end{equation}
\begin{equation}\delta_Z=\cot\theta_W\hskip 2pt f^{(6)}_W
\hskip 2pt\frac{M^2_Z}{2\Lambda^2};
\quad x_Z=-\tan\theta_W\hskip 2pt x_\gamma;
\quad y_Z=\cot\theta_W\hskip 2pt y_\gamma.\label{xgamma}\end{equation}
According to (\ref{deltaz}) and (\ref{xgamma}), in this model there are only
three independent couplings which we can choose to be $x_\gamma$, $y_\gamma$
and $\delta_Z$.\footnote{As mentioned in \cite{zeppenfeld}, the correlations
between different anomalous trilinear gauge boson couplings exhibited
in Eqs.(\ref{deltaz}) and (\ref{xgamma}) are due to the truncation of the
effective Lagrangian (\ref{lagra1}) at the dimension 6 level, and do not
hold any longer when dimension 8 (or higher) operators are included. It is
interesting to notice that the relation between $x_\gamma$ and
$x_Z$ in (\ref{xgamma}) was first introduced in \cite{kuroda} on the basis of
global $SU(2)_W$ symmetry for W dynamics and $W_3-\gamma$ mixing.}
Of these, $x_\gamma$ and $y_\gamma$ are directly bound
from Table 2, and the constraints on $x_Z$ and $y_Z$ are simply obtained
from the previous ones using last two relations of Eq.(\ref{xgamma}).
Finally, the bound on $\delta_Z$ is obtained by combining that on $x_Z$
with Eq.(\ref{betall}). This procedure gives the tightest bounds on
$\delta_Z$: the other ones, utilizing the inequalities (\ref{betatl}) or
(\ref{betatt}) would be less stringent. This is due to the fact that
the regions allowed by the $W^+_LW^-_L$ production cross sections are
much more restricted than those determined by the other final polarizations,
as can be seen by comparing Figs. 3 to 5. Numerically, we find the values
reported in Table 3, to be compared with the model independent ones in Table 2.
\begin{table}
\centering
\begin{tabular}{|c|c|c|c|c|c|}
\hline
$E_{CM}(TeV)$ & $x_\gamma\hskip2pt (10^{-3})$ & $y_\gamma\hskip2pt (10^{-3})$
& $\delta_Z\hskip2pt (10^{-3})$ & $x_Z\hskip2pt (10^{-3})$
& $y_Z\hskip2pt (10^{-3})$ \\ \hline
$0.5$ & $-1.8\div 1.8$ & $-8.6\div 9.2$ &$-3.7\div 3.7$ & $-1.0\div 1.0$
& $-16\div 17$ \\ \hline
$1$ & $-0.5\div 0.5$ & $-3.0\div 3.0$ &$-1.0\div 1.0$ & $-0.3\div 0.3$
& $-5.5\div 5.5$ \\ \hline
\end{tabular}
\caption{95\% CL limits for the model with three independent anomalous
couplings. Same inputs as in Table 1.} \label{tab:tab3} \end{table}
\par In conclusion, summarizing the previous analysis, the
results obtained show the potential of the approach to derive
bounds on the anomalous trilinear boson couplings, based on cross sections
integrated with suitably defined cuts and combinations of all possible
initial and final polarizations. This allows to separately constrain the
$CP$ conserving couplings in a model independent way with high sensitivity,
typically of the order of $10^{-3}-10^{-2}$ at $E_{CM}=0.5\hskip 3pt TeV$.
Particularly stringent bounds can be expected for dynamical models beyond the
SM with reduced number of independent couplings.\par
In principle, one could include in this kind of analysis also the anomalous
coupling $\delta_\gamma$, still $CP$ conserving, which would be induced
{\it e.g.} by a dimension 8 contribution to (\ref{lagra1})
\cite{gounaris3}.
Having, in this case, equal numbers of polarized observables and anomalous
couplings, separate constraints could still be found.\par
The bounds derived above are approaching the order of magnitude of the
radiative corrections to the SM couplings \cite{fleischer}. Thus, the next
step should be the combination in the fitting procedure of the SM radiative
corrections with the anomalous gauge boson couplings.
\vspace{4cm}
\begin{center}{\bf Acknowledgements}\end{center}
\noindent
We are grateful to Dr. V. V. Andreev for stimulating discussions.
One of us (AAP) acknowledges the support and the hospitality
of INFN-Sezione di Trieste and of the International Centre for
Theoretical Physics, Trieste.
The work of (NP) has been supported in part by the
Human Capital and Mobility Programme, EEC Contract ERBCHRXCT930132.

\newpage
\section*{Figure captions}
\begin{description}

\item{\bf Fig.1}
$x_{\gamma}$-dependence of the $\chi^2$ in Eq.(\ref{chisquare}) for
$e^+e^-\to W^+W^-$ at $E_{CM}=500\hskip 3pt GeV$, integrated luminosity
$L_{int}= 20\hskip 2pt fb^{-1}$. `unpol' and `unpol-opt' refer to the
unpolarized cross section integrated over the angular range $\vert z\vert
<0.98$ and over the `optimal' kinematical region ($-0.98\div 0.0$),
respectively.
`$LR$' and `$RL$' refer to polarized cross sections integrated up to
$z_{opt}=-0.2$ and $z_{opt}=0.7$, respectively.

\item{\bf Fig.2} Allowed domains (95\% C.L.) from $e^-e^+\to W^-_LW^+_L$
with polarized ($RL$, $LR$) and unpolarized initial beams
at $E_{CM}=0.5\hskip 3pt TeV$, inputs as specified in Table 1.

\item{\bf Fig.3} Same as Fig.2, magnified allowed domain around the origin,
and combined area allowed by $RL$ and $LR$ cross sections. The smaller area
around origin refers to $E_{CM}=1\hskip 3pt TeV$,
$L_{int}=50\hskip 2pt fb^{-1}$.

\item{\bf Fig.4} Allowed domains (95\% C.L.) for
($x_{\gamma}+y_{\gamma},\hskip 2pt x_Z+y_Z+2\delta_Z$)
from $e^-e^+\to W^-_LW^+_T+W^-_TW^+_L$ with same inputs as in Fig.2 and Fig.3.

\item{\bf Fig.5} Allowed domains (95\% C.L.) for
($y_{\gamma},\hskip 2pt y_Z+\delta_Z\hskip 2pt \frac{1-\beta_W^2}{2}$)
from $e^-e^+\to W^-_TW^+_T$ with same inputs as in Fig.2 and Fig.3.

\end{description}

\end{document}